# Strong lateral exchange coupling and current-induced switching in single-layer ferrimagnetic films with patterned compensation temperature

Zhentao Liu[1,2], Zhaochu Luo[1,2,3,4,*], Ivan Shorubalko[5], Christof Vockenhuber[6], Laura J. Heyderman[1,2], Pietro Gambardella[7,**], Aleš Hrabec[1,2,7,***]

[1]Laboratory for Mesoscopic Systems, Department of Materials, ETH Zurich, 8093 Zurich, Switzerland

[2]Laboratory for Multiscale Materials Experiments, Paul Scherrer Institute, 5232 Villigen PSI, Switzerland

[3]State Key Laboratory for Mesoscopic Physics, School of Physics, Peking University, 100871 Beijing, People's Republic of China

[4]Beijing Key Laboratory for Magnetoelectric Materials and Devices, 100871 Beijing, People's Republic of China

[5]Transport at Nanoscale Interfaces Laboratory, Empa - Swiss Federal Laboratories for Materials Science and Technology, 8600 Dübendorf, Switzerland

[6]Laboratory of Ion Beam Physics, ETH Zürich, 8093 Zürich, Switzerland

[7]Laboratory for Magnetism and Interface Physics, Department of Materials, ETH Zurich, 8093 Zurich, Switzerland

**Strong, adjustable magnetic couplings are of great importance to all devices based on magnetic materials. Controlling the coupling between adjacent regions of a single magnetic layer, however, is challenging. In this work, we demonstrate strong exchange-based coupling between arbitrarily shaped regions of a single ferrimagnetic layer. This is achieved by spatially patterning the compensation temperature of the ferrimagnet by either oxidation or He$^+$ irradiation. The coupling originates at the lateral interface between regions with different compensation temperature and scales inversely with their width. We show that this coupling generates large lateral exchange coupling fields and we demonstrate its application to control the switching of magnetically compensated dots with an electric current.**

In spintronic architectures based on magnetic multilayers [1-3], interlayer couplings such as the Ruderman–Kittel–Kasuya–Yosida interaction [4,5], exchange coupling leading to exchange bias [6,7], and the dipolar interaction [8-10] allow for the tuning of the magnetic stability and electrical properties of the device [1-3]. The exchange interaction, composed of symmetric and antisymmetric parts, provides the strongest coupling channel in magnetic systems. The symmetric part favors collinear magnetic configurations and is commonly exploited in multilayers to provide direct exchange coupling between, for example, two ferromagnets [11] or a ferromagnet and an antiferromagnet [6], and indirect coupling between two ferromagnets separated by a nonmagnetic spacer [4,5]. The antisymmetric part, known as the Dzyaloshinskii–Moriya interaction (DMI), favors non-collinear magnetic textures, but is generally indirect and weaker in multilayer systems [12-14]. Controlling the coupling between adjacent regions of a single magnetic layer is more challenging [15-17]. In single layers, the interfacial DMI provides a means to couple planar structures [16,18], which has enabled the realization of electrically-controlled magnetic logic devices [19-23]. However, the strength of this coupling is limited by the interface properties [24]. The stronger collinear exchange coupling in magnetic multilayers thus lacks a counterpart in the lateral direction.

* zhaochu.luo@pku.edu.cn
**pietro.gambardella@mat.ethz.ch
***ales.hrabec@psi.ch

In this work, we realize a strong lateral coupling based on the exchange interaction in a single magnetic layer. We take inspiration from an approach previously developed for synthetic ferrimagnetic systems consisting of stacked layers of rare-earth transition-metal ferrimagnets with different magnetic compensation temperature ($T_\mathrm{M}$) [25-31]. This type of multilayer is also known as an exchange spring magnet and includes a compensation wall. We further develop this method, applying it to a single-layer ferrimagnetic alloy. In such an alloy, the strong intra-lattice coupling between the transition-metal atoms and the weaker inter-lattice coupling between the rare-earth and transition metal atoms can be separately tuned by altering the composition [32] or microstructure [33], and through reduction/oxidation reactions [34,35]. Recently, He$^+$ irradiation has been used to modify $T_\mathrm{M}$ in a Co/Tb multilayer in order to create multidomain configurations with dimensions of several μm [30,36]. Here we show that patterning of $T_\mathrm{M}$ in a single GdCo film by either selective oxidation or He$^+$ irradiation leads to strong lateral exchange coupling between regions with different $T_\mathrm{M}$. We show how the coupling varies as a function of temperature and width of the patterned regions. We discuss the strength of the coupling and compare the exchange interaction in our planar structures with that found in multilayer systems. We further combine spin-orbit torques [3] and lateral coupling in a Pt|GdCo bilayer to demonstrate selective current-induced switching of adjacent magnetic domains, which results in reproducible manipulation of lateral exchange bias.

Our Ta(1 nm)|Pt(5 nm)|Gd$_{0.3}$Co$_{0.7}$($x$ nm)|Ta(2 nm) multilayers possess out-of-plane (OOP) magnetization with $x$ in the range from 3.2 to 6.2 nm, where $T_\mathrm{M}$ can be tuned by changing the stoichiometric ratio or the thickness of the GdCo layer as shown in the Supplemental Material [37] (see, also, reference [38] therein). To spatially modify $T_\mathrm{M}$, we use an oxygen plasma to partially oxidize the magnetic film in specific regions. The effect of oxidation is verified by Rutherford Backscattering technique [37]. In particular, a GdCo film with $T_\mathrm{M} = T_\mathrm{c2}$ above room temperature (RT) can be oxidized in order to lower the compensation point below room temperature ($T_\mathrm{c1}$). Magneto-optic Kerr effect (MOKE) measurements, which are predominantly sensitive to the magnetization of the Co sublattice, show a reversal in the hysteresis loop after oxidation of a GdCo film [37], indicating that $T_\mathrm{M}$ is suppressed below room temperature after the oxidation.

For a device containing two regions with different $T_\mathrm{M}$ ($T_\mathrm{c2} > T_\mathrm{c1}$), one can distinguish between three temperature scenarios illustrated in Fig. 1(a-c): (i) the temperature is higher than both compensation temperatures ($T > T_{c2}$), (ii) the temperature is lower than both compensation temperatures ($T < T_{c1}$), and (iii) the temperature is lower than the compensation temperature of the original film but higher than the compensation temperature of the partially oxidized film ($T_{c2} > T > T_{c1}$).

In the temperature scenarios with $T > T_{c2}$ and $T < T_{c1}$ [Fig. 1(a) and (b)], all of the Co magnetic moments are parallel to each other and all of the Gd moments are antiparallel to the Co moments, minimizing the exchange and magnetic anisotropy energy. The net magnetization is then given by the sum of the two sublattice magnetizations, $\boldsymbol{M}_{\mathbf{net}} = \boldsymbol{M}_{\mathrm{Co}} + \boldsymbol{M}_{\mathrm{Gd}}$. At a temperature $T > T_{c2}$ ($T < T_{c1}$), $\boldsymbol{M}_{\mathbf{net}}$ is parallel to $\boldsymbol{M}_{\mathrm{Co}}(\boldsymbol{M}_{\mathrm{Gd}})$ in both the pristine and oxidized regions of the film [Fig. 1(a,b)]. Because the neighboring Co moments are strongly exchange-coupled and prefer to maintain a parallel alignment, not only within the two different regions but also across the interface between them, the net magnetization in the two regions is effectively (trivially) ferromagnetically coupled.

In the temperature range $T_{c2} > T > T_{c1}$, however, $\boldsymbol{M}_{\mathbf{net}}$ is parallel to $\boldsymbol{M}_{\mathrm{Co}}$ in the region with $T_{c1}$ but parallel to $\boldsymbol{M}_{\mathrm{Gd}}$ in the region with $T_{c2}$. Hence, the low-energy configuration that minimizes the exchange energy between the Co moments across the oxidation interface is an antiparallel state of the net magnetization [Fig. 1(c)]. At the same time, the dipolar energy is reduced since the net magnetization of left-hand and right-hand regions form a flux-closure configuration. A sufficiently high external magnetic field can twist the antiparallel magnetization state to the parallel state [Fig. 1(d)], so reducing the Zeeman energy [26,27,30,36,39]. This switching process is accompanied by the creation of a DW for the Co and Gd moments associated with an energy cost $E_{\mathrm{DW}}$. In the regime where the dipolar energy is negligible (see Supplemental Material [37]), the DW energy determines the strength of the effective antiparallel coupling $J_{\mathrm{AP}}$ between the net magnetization in the regions with compensation temperatures of $T_{c1}$ and $T_{c2}$. The antiparallel coupling gives rise to an effective exchange coupling field $H_{\mathrm{EC}}$:

$$\mu_0 H_{\mathrm{EC}} = \frac{J_{\mathrm{AP}}}{M_{\mathrm{net}}} \cong \frac{\lambda_{\mathrm{DW}}}{M_{\mathrm{net}}w} = \frac{4\sqrt{A_{\mathrm{eff}}K_{\mathrm{eff}}} - \pi D}{M_{\mathrm{net}}w}, \qquad (1)$$

where $\lambda_{\mathrm{DW}}, A_{\mathrm{eff}}, K_{\mathrm{eff}}, D, M_{\mathrm{net}}$ and $w$ are the DW energy density, effective exchange stiffness, effective magnetic anisotropy, DMI strength, net magnetization and the width of the switched area, respectively [37].

The impact of the coupling on $\boldsymbol{M}_{\mathrm{Co}}$ and $\boldsymbol{M}_{\mathbf{net}}$ [illustrated in Fig. 1(c)] can be demonstrated by selectively oxidizing a checkerboard pattern with square width of 800 nm in a film with $T_{\mathrm{M}} >$ RT, as schematically shown in the inset of Fig. 1(e). After removal of a large magnetic field saturating the sample, the Kerr contrast, predominantly arising from the Co sublattice, displays a uniform state [Fig. 1(e)]. In contrast, the magnetic force microscopy image, where the stray fields produced by the net magnetization are detected, reveals an alternating contrast [Fig. 1(f)]. The nanoscale magnetization pattern is predominantly driven by lateral exchange coupling whereas, increasing the dimensions towards a micrometer scale pattern would lead to an increase in the influence of the dipolar interaction [36]. To further demonstrate this lateral exchange coupling at ambient temperature, we patterned a 50 μm × 50 μm square with half of the square being oxidized [light and dark grey regions of the square in Fig. 1(g)]. The as-grown part of the square [white region of the square in Fig. 1(g)] is compensated with $T_M$ slightly above RT, such that its magnetization is negligible. The oxidized region has its $T_M$ far below RT, resulting in a lateral exchange-biased structure. As shown in Fig. 1(g), by warming up from 250 K to 300 K in an applied magnetic field $\mu_0 H_z = \pm 6$ T in order to preset the state of the compensated region, a switching of the exchange-biased hysteresis loop ($\mu_0 H_{EB} = \pm 24$ mT) can be observed depending on the state of the compensated region [37].

In order to confirm the interfacial origin of the exchange coupling, we selectively oxidized tracks with widths in the range from 50 to 200 nm in the original GdCo films. The electric detection of the magnetic state ($\boldsymbol{M}_{Co}$) is performed via 1-μm-wide Hall bars [Fig. 2(a)]. Full and minor hysteresis loops at temperatures ranging from 300 K down to 200 K are then recorded [37]. An example set of hysteresis loops for a 150 nm-wide track measured at 300, 220 and 200 K, corresponding to the three distinct temperature ranges, is presented in Fig. 2(a). As expected, the hysteresis loops in the temperature range with $T > T_{c2}$ [300 K loop in Fig. 2(a)] and $T < T_{c1}$ [200 K loop in Fig. 2(a)] are trivial since the net magnetization of both the $T_{c1}$ and $T_{c2}$ regions simply switch when applying a sufficient magnetic field. In the temperature range $T_{c2} > T > T_{c1}$, on application of a large enough magnetic field, the Zeeman energy will cause the net magnetization of both regions to follow the applied field, leading to $\boldsymbol{M}_{Co}$ in the oxidized and non-oxidized regions pointing in opposite directions. On reducing the magnetic field, the exchange coupling overcomes the Zeeman interaction resulting in parallel orientation of the two Co magnetic sublattices. This is accompanied by an enhancement of the Hall signal [37]. After surpassing the coercive field of the surrounding non-oxidized GdCo layer, the magnetization switches while maintaining the parallel Co magnetic configuration. When the field is further increased, the net magnetization in the two regions again aligns in parallel.

The systematic measurement of the exchange coupling strength at different temperatures is summarized in Fig. 2(b). In line with the proposed mechanism, no exchange coupling field is observed when the temperature is above or below both $T_{c1}$ and $T_{c2}$. However, once the temperature is below $T_M$ of the surrounding non-oxidized GdCo layer, an increase in the exchange coupling field can be observed as the magnetization of the track approaches its compensation point on reducing the temperature. The increase in the exchange coupling field is caused by the reduction of the net magnetization of the tracks, which can be quantitatively described by Equation (1) and fitted to the experimental data. By reducing the track width from 200 to 50 nm, the exchange coupling field strength is further increased. This confirms the interfacial origin of the coupling effect, which becomes stronger in devices with reduced lateral dimensions. The exchange coupling fields reach values as high as 2.5 T. It should be noted that, in contrast to the lateral exchange coupling, the dipolar coupling mechanism reported previously [30,36] decreases in smaller devices and is therefore not useful for miniaturization of devices. The micromagnetic simulations of the effective coupling field with and without dipolar field is shown in the Supplemental Material [37] (see, also, references [40,41] therein).

To provide microscopic insight into the switching mechanism, a GdCo film with $T_M$ above RT is patterned into 40 μm long tracks of various width. Imaging in a wide field polar Kerr microscope reveals that the magnetization reversal is driven by DW propagation along the track [Fig. 3(a)]. The reverse domains [given by white contrast in Fig. 3(a)] are created at the sharp tips at both ends of the track and propagate towards the center. Moreover, the curvature of the moving DWs suggests that the DW is strongly dragged by the lateral interface. The curvature angle of the DW respect to the DW moving direction is in stark contrast to the case where the DW motion is hindered by pinning at the edges of a magnetic racetrack [42]. The DW can then serve as a probe of the magnetic coupling [18]. We measure the so-called stopping field at which the DW motion is arrested, meaning that the exchange coupling is balanced by the Zeeman energy. The stopping field can be deduced from the hysteresis loop, an example of which is shown in Fig. 3(a), and corresponds to the field value where the Kerr rotation switches sign. A clear increase of the stopping field is observed when the track width is reduced from 800 to 200 nm [Fig. 3(b)], which again indicates the interfacial origin of the coupling. In addition, the stopping fields are more than one order of magnitude higher than the ones originating from the DMI-driven chiral coupling mechanism, which opens routes to more efficient, tunable lateral couplings [18].

In addition to using oxidation, we can realize lateral exchange coupling using He$^+$ irradiation, where the spatial modification of $T_M$ is achieved by introducing defects and oxygen atoms into the film [30,31,36,43]. The He$^+$ irradiation technique offers a good alternative to patterning using oxidation, with exquisite control over the desired coupling strength by changing the irradiation dose and a smaller achievable feature size of 5 nm [44] (see Supplemental Material [37] for the relevant experimental results).

To illustrate the potential of the lateral coupling in a single ferrimagnetic film for applications, we designed a planar counterpart of exchange-biased ferromagnet/antiferromagnet bilayers, in which the exchange bias is manipulated via the spin-orbit torques (SOTs) [45]. We have patterned a compensated GdCo film ($T_M \geq \text{RT}$) into cross-like structures placed on a Pt conduit. The Pt layer is used as a source of sizeable DMI as well of spin current [3,19]. Each cross is divided into five squares, where the four surrounding squares are selectively oxidized, whereas the central square remains in its pristine state [Fig. 4(a)]. By applying a large ($\pm 250$ mT) OOP magnetic field, only the oxidized regions can be switched since they have a sizeable net magnetization [Fig. 4(c)]. In order to utilize SOTs to switch the magnetization, an IP external magnetic field ($H$) is applied along the current direction ($J$) [3]. Starting from the case where $\boldsymbol{M}_{Co}$ is parallel across the entire device, a series of current pulses causes the magnetization in the entire structure to be switched [Fig. 4(d)]. This is caused by the SOT-driven switching of the uncompensated squares, and the central square switches with them due to the strong lateral exchange coupling. From the hysteresis loops in Fig. 4(g), we then see a clear exchange bias field of $\pm 30$ mT whose polarity depends on the orientation of the compensated square set by the SOT. The electric switching of lateral exchange bias is highly reproducible (see Supplemental Material [37]). The unique combination of SOT switching and lateral exchange coupling therefore provides a means to achieve magnetic states, which are otherwise only accessible via a field cooling protocol [Fig. 1(g)]. To corroborate the proposed mechanism, we have also fabricated a complementary cross-structure where an oxidized square is placed in the center of the cross while the outer four squares are non-oxidized [Fig. 4(b)]. While a large magnetic field can be used to switch the magnetization of the inner square only [Fig. 4(e)], the SOT is not able to induce switching of the central region because the lateral exchange coupling to the surrounding regions is too strong [Fig. 4(f)].

In conclusion, the lateral exchange coupling reported in single-layer ferrimagnetic devices with sub-micron dimensions provides an important addition to the family of intra-layer couplings. Unlike in vertical stacks, where the interfacial exchange coupling always contains an immobile compensation wall [25-29], the lateral interfacial exchange coupling strength can be easily tuned by modifying the device geometry, altering the $He^+$ irradiation dose, changing the oxidation exposure, or changing the temperature. The coupling is given by the local exchange interaction between the transition metal atoms across the interface between regions with different compensation temperatures, which allows for device downscaling without the loss of the coupling strength. The estimated strength of the lateral interfacial exchange coupling is much larger than the volume-like dipolar coupling in ferrimagnetic multilayers [30,36] and it is one order of magnitude stronger than the DMI-mediated coupling in single-layer ferromagnets [16,18]. Furthermore, by combining the interfacial exchange coupling with current driven SOT switching, we are able to access both magnetization states of a compensated ferrimagnet, which can be otherwise only be accessed by a field-cooling protocol. The lateral exchange coupling, where the coupling strength is maintained on downscaling of the device, serves as an important counterpart to the coupling in vertical devices and opens up the possibility for new functionalities in planar devices.

**Acknowledgements:**

This project received funding from the Swiss National Science Foundation (Grant Agreements No. 200021_182013 and 200020_200465). A.H. was funded by the European Union's Horizon 2020 research and innovation program under Marie Skłodowska-Curie grant agreement No. 794207 (ASIQS). Z.L. and L.J.H. acknowledge funding from the European Union's Horizon 2020 FET-Open program under Grant Agreement No. 861618 (SpinEngine). Z.L. also acknowledge funding from the National Natural Science Foundation of China (No. 52271160). We thank Max Doebeli for help with the analysis of the ERDA measurements.

**Data Availability:**

The data that support this study are available via the Zenodo repository, 10.5281/zenodo.6936908, Ref. [46].

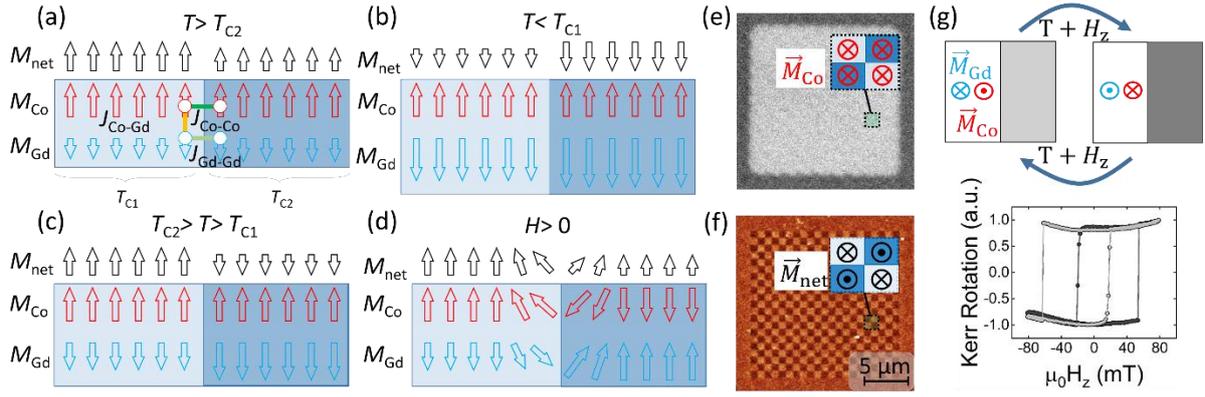

**FIG.1 Demonstration of the lateral exchange coupling principle.** (a-d) Schematic of interfacial exchange coupling at different temperatures. $M_{net} = M_{Co} + M_{Gd}$. The dark and light blue background correspond to the non-oxidized and oxidized regions, respectively. The $J_{Co-Co}$ (green), $J_{Gd-Gd}$ (light green) and $J_{Co-Gd}$ (yellow) represent exchange coupling between different elements, respectively. (e) Kerr image and (f) MFM image of the checkboard pattern spontaneously formed in 800-nm-wide GdCo squares at zero field. The symbols in the insets represent the orientation of the magnetization of the Co sublattice (red) and the net magnetization (black) while the dark and light blue background correspond to the non-oxidized and oxidized sections, respectively. (g) Hysteresis loops measured on a 50 μm × 50 μm GdCo square with one half being magnetically compensated (in white) and the other half being oxidized (in light or dark gray). The dark and light gray data points correspond to measurements taken after field cooling the sample from $T$=250 K to 300 K at −6 T and +6 T ($H_z$), respectively.

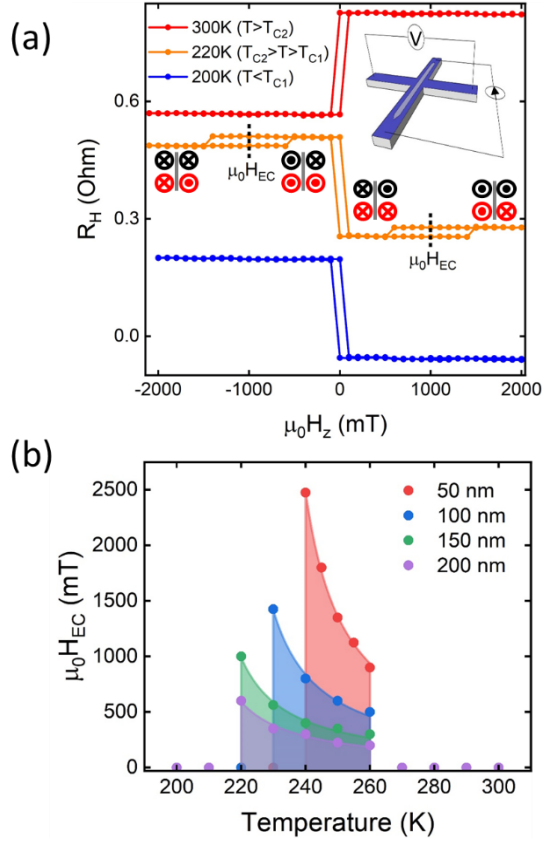

**FIG.2 Temperature dependence of the lateral exchange coupling.** (a) A set of hysteresis loops measured via Hall resistance as a function of applied magnetic field of a 150-nm-wide track at different temperatures. The plots are shifted by 0.3 and 0.6 Ohm (for 220 and 300 K) for clarity. In the inset is a schematic of the oxidized track (in gray) on a 1-μm-wide Hall bar (in blue) used for anomalous Hall resistance measurements. The relevant magnetization states of oxidized|non-oxidized regions are depicted (black for $M_{net}$ and red for $M_{Co}$). (b) Exchange coupling field versus temperature plots for different track widths in the range 50 to 200 nm. The lines are fits according to Eq. (1). The thickness of the GdCo film is 4.6 nm. Note that the data points outside the shaded regions, where the exchange coupling field is null, overlap.

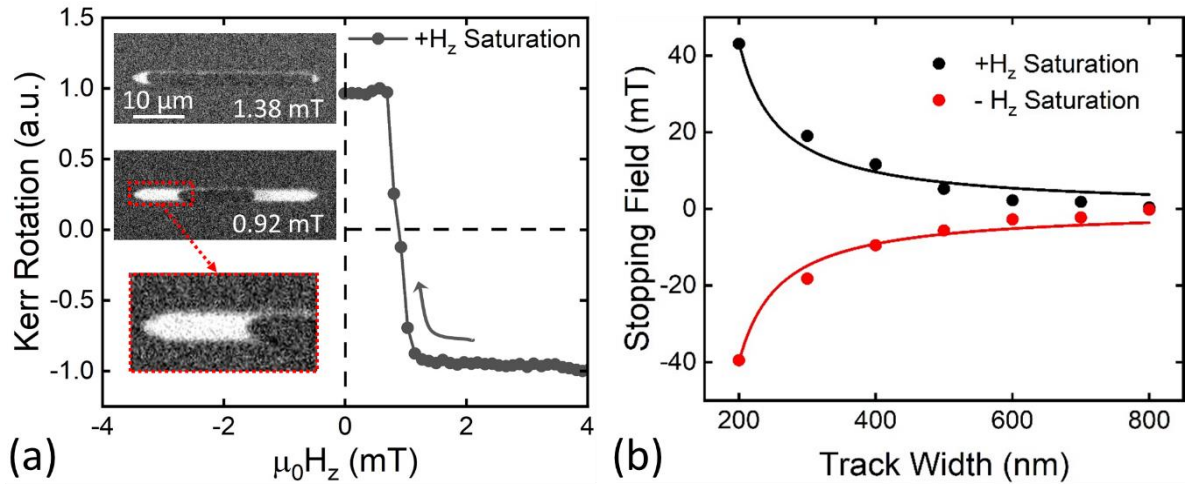

**FIG.3 Determination of the stopping field associated with the lateral exchange coupling.** (a) Kerr contrast as a function of decreasing magnetic field obtained from a 2-μm-wide track. The thickness of GdCo is 6.5 nm. Examples of Kerr micrographs at different fields and enlargement of DW profile are displayed in the inset. (b) Stopping field *versus* track width in GdCo patterned by oxidation. The lines in (b) are fits according to Eq. (1).

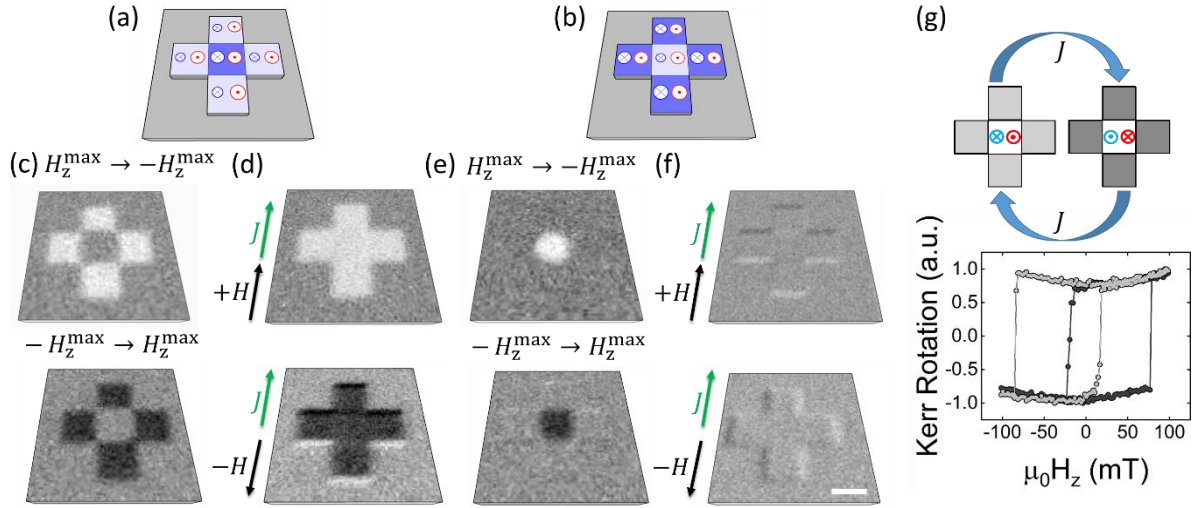

**FIG.4 Electric control of lateral exchange bias.** (a-b) Schematic of a cross structure divided into four oxidized squares surrounding a central compensated square (a) and vice versa (b). The symbols depict the magnetization of the Co (red) and Gd (blue) sublattices, respectively. (c,e) Kerr differential images after applying a $\pm 250$ mT OOP magnetic field. (d,f) Kerr images after applying $100 \times 50$ ns current pulses at $J = 1.35 \times 10^{11}$ A/m², showing that the entire cross has switched from its initial state (d) and no switching (f). An in-plane magnetic field of $H = \pm 200$ mT is applied along the current direction. (g) Hysteresis loops of the four outer squares depicted in (a) for the two magnetic orientations of the compensated central square, **M**(Co) = ⊙ (light gray) and **M**(Co) = ⊗ (dark gray). The scale bars is 2 µm.

# Reference:

[1] C. Chappert, A. Fert, and F. N. Van Dau, in *Nanoscience And Technology: A Collection of Reviews from Nature Journals* (World Scientific, 2010), pp. 147-157.

[2] S. Parkin and S.-H. Yang, Memory on the racetrack, Nature nanotechnology **10**, 195 (2015).

[3] A. Manchon, J. Železný, I. M. Miron, T. Jungwirth, J. Sinova, A. Thiaville, K. Garello, and P. Gambardella, Current-induced spin-orbit torques in ferromagnetic and antiferromagnetic systems, Reviews of Modern Physics **91**, 035004 (2019).

[4] P. Grünberg, R. Schreiber, Y. Pang, M. Brodsky, and H. Sowers, Layered magnetic structures: Evidence for antiferromagnetic coupling of Fe layers across Cr interlayers, Physical review letters **57**, 2442 (1986).

[5] S. Parkin, R. Bhadra, and K. Roche, Oscillatory magnetic exchange coupling through thin copper layers, Physical Review Letters **66**, 2152 (1991).

[6] J. Nogués and I. K. Schuller, Exchange bias, Journal of Magnetism and Magnetic Materials **192**, 203 (1999).

[7] R. Stamps, Mechanisms for exchange bias, Journal of Physics D: Applied Physics **33**, R247 (2000).

[8] E. Hill, S. Tomlinson, and J. Li, The role of dipole coupling in multilayers, Journal of applied physics **73**, 5978 (1993).

[9] R. V. Chopdekar, B. Li, T. A. Wynn, M. S. Lee, Y. Jia, Z. Liu, M. D. Biegalski, S. T. Retterer, A. T. Young, and A. Scholl, Nanostructured complex oxides as a route towards thermal behavior in artificial spin ice systems, Physical Review Materials **1**, 024401 (2017).

[10] D. Y. Sasaki, R. V. Chopdekar, S. T. Retterer, D. Y. Jiang, J. K. Mason, M. S. Lee, and Y. Takamura, Formation of Complex Spin Textures in Thermally Demagnetized La 0.7 Sr 0.3 Mn O 3 Artificial-Spin-Ice Structures, Physical Review Applied **17**, 064057 (2022).

[11] E. E. Fullerton, J. Jiang, and S. Bader, Hard/soft magnetic heterostructures: model exchange-spring magnets, Journal of Magnetism and Magnetic Materials **200**, 392 (1999).

[12] D.-S. Han, K. Lee, J.-P. Hanke, Y. Mokrousov, K.-W. Kim, W. Yoo, Y. L. Van Hees, T.-W. Kim, R. Lavrijsen, and C.-Y. You, Long-range chiral exchange interaction in synthetic antiferromagnets, Nature materials **18**, 703 (2019).

[13] A. Fernández-Pacheco, E. Vedmedenko, F. Ummelen, R. Mansell, D. Petit, and R. P. Cowburn, Symmetry-breaking interlayer Dzyaloshinskii–Moriya interactions in synthetic antiferromagnets, Nature materials **18**, 679 (2019).

[14] C. O. Avci, C.-H. Lambert, G. Sala, and P. Gambardella, Chiral coupling between magnetic layers with orthogonal magnetization, Physical review letters **127**, 167202 (2021).

[15] S. H. Skjærvø, C. H. Marrows, R. L. Stamps, and L. J. Heyderman, Advances in artificial spin ice, Nature Reviews Physics **2**, 13 (2020).

[16] Z. Luo, T. P. Dao, A. Hrabec, J. Vijayakumar, A. Kleibert, M. Baumgartner, E. Kirk, J. Cui, T. Savchenko, and G. Krishnaswamy, Chirally coupled nanomagnets, Science **363**, 1435 (2019).

[17] A. Hrabec, Z. Luo, L. J. Heyderman, and P. Gambardella, Synthetic chiral magnets promoted by the Dzyaloshinskii–Moriya interaction, Applied Physics Letters **117**, 130503 (2020).

[18] Z. Liu, Z. Luo, S. Rohart, L. J. Heyderman, P. Gambardella, and A. Hrabec, Engineering of Intrinsic Chiral Torques in Magnetic Thin Films Based on the Dzyaloshinskii-Moriya Interaction, Physical Review Applied **16**, 054049 (2021).

[19] Z. Luo, A. Hrabec, T. P. Dao, G. Sala, S. Finizio, J. Feng, S. Mayr, J. Raabe, P. Gambardella, and L. J. Heyderman, Current-driven magnetic domain-wall logic, Nature **579**, 214 (2020).

[20] Z. Luo, S. Schären, A. Hrabec, T. P. Dao, G. Sala, S. Finizio, J. Feng, S. Mayr, J. Raabe, and P. Gambardella, Field-and current-driven magnetic domain-wall inverter and diode, Physical Review Applied **15**, 034077 (2021).

[21] Z. Zeng, Z. Luo, L. J. Heyderman, J.-V. Kim, and A. Hrabec, Synchronization of chiral vortex nano-oscillators, Applied Physics Letters **118**, 222405 (2021).

[22] T. P. Dao, M. Müller, Z. Luo, M. Baumgartner, A. Hrabec, L. J. Heyderman, and P. Gambardella, Chiral domain wall injector driven by spin–orbit torques, Nano Letters **19**, 5930 (2019).
[23] F. Ummelen, H. Swagten, and B. Koopmans, Racetrack memory based on in-plane-field controlled domain-wall pinning, Scientific reports **7**, 1 (2017).
[24] A. Belabbes, G. Bihlmayer, F. Bechstedt, S. Blügel, and A. Manchon, Hund's rule-driven dzyaloshinskii-moriya interaction at 3 d− 5 d interfaces, Physical review letters **117**, 247202 (2016).
[25] B. Hebler, P. Reinhardt, G. Katona, O. Hellwig, and M. Albrecht, Double exchange bias in ferrimagnetic heterostructures, Physical Review B **95**, 104410 (2017).
[26] P. Hansen, New type of compensation wall in ferrimagnetic double layers, Applied physics letters **55**, 200 (1989).
[27] T. Kobayashi, H. Tsuji, S. Tsunashima, and S. Uchiyama, Magnetization process of exchange-coupled ferrimagnetic double-layered films, Japanese Journal of Applied Physics **20**, 2089 (1981).
[28] C. Blanco-Roldán, Y. Choi, C. Quiros, S. Valvidares, R. Zarate, M. Vélez, J. Alameda, D. Haskel, and J. I. Martin, Tuning interfacial domain walls in GdCo/Gd/GdCo′ spring magnets, Physical Review B **92**, 224433 (2015).
[29] F. Stobiecki, T. Atmono, S. Becker, H. Rohrmann, and K. Röll, Investigation of interface wall energy σw and coercivity HC in exchange-coupled double layers (ECDLs), Journal of magnetism and magnetic materials **148**, 497 (1995).
[30] Ł. Frąckowiak, F. Stobiecki, G. D. Chaves-O'Flynn, M. Urbaniak, M. Schmidt, M. Matczak, A. Maziewski, M. Reginka, A. Ehresmann, and P. Kuświk, Subsystem domination influence on magnetization reversal in designed magnetic patterns in ferrimagnetic Tb/Co multilayers, Scientific Reports **11**, 1 (2021).
[31] M. Krupinski, J. Hintermayr, P. Sobieszczyk, and M. Albrecht, Control of magnetic properties in ferrimagnetic GdFe and TbFe thin films by He+ and Ne+ irradiation, Physical Review Materials **5**, 024405 (2021).
[32] K. Buschow, Intermetallic compounds of rare-earth and 3d transition metals, Reports on Progress in Physics **40**, 1179 (1977).
[33] D.-H. Kim, M. Haruta, H.-W. Ko, G. Go, H.-J. Park, T. Nishimura, D.-Y. Kim, T. Okuno, Y. Hirata, and Y. Futakawa, Bulk Dzyaloshinskii–Moriya interaction in amorphous ferrimagnetic alloys, Nature materials **18**, 685 (2019).
[34] M. Huang, M. U. Hasan, K. Klyukin, D. Zhang, D. Lyu, P. Gargiani, M. Valvidares, S. Sheffels, A. Churikova, and F. Büttner, Voltage control of ferrimagnetic order and voltage-assisted writing of ferrimagnetic spin textures, Nature Nanotechnology **16**, 981 (2021).
[35] E. Kirk, C. Bull, S. Finizio, H. Sepehri-Amin, S. Wintz, A. K. Suszka, N. S. Bingham, P. Warnicke, K. Hono, and P. Nutter, Anisotropy-induced spin reorientation in chemically modulated amorphous ferrimagnetic films, Physical Review Materials **4**, 074403 (2020).
[36] Ł. Frąckowiak, P. Kuświk, G. D. Chaves-O'Flynn, M. Urbaniak, M. Matczak, P. P. Michałowski, A. Maziewski, M. Reginka, A. Ehresmann, and F. Stobiecki, Magnetic domains without domain walls: A unique effect of He+ Ion bombardment in ferrimagnetic Tb/Co films, Physical Review Letters **124**, 047203 (2020).
[37] See Supplemental Material at [url] for additional information of sample fabrication, sample thickness and oxidation dependence, transport measurement, Hall measurement, FIB irradiation and micromagnetic simulation details., (2022).
[38] R. Malmhäll and T. Chen, Thickness dependence of magnetic hysteretic properties of rf‐sputtered amorphous Tb‐Fe alloy thin films, Journal of Applied Physics **53**, 7843 (1982).
[39] A. Hrabec, N. Nam, S. Pizzini, and L. Ranno, Magnetization reversal in composition-controlled Gd1−x Co x ferrimagnetic films close to compensation composition, Applied Physics Letters **99**, 052507 (2011).
[40] A. Vansteenkiste, J. Leliaert, M. Dvornik, M. Helsen, F. Garcia-Sanchez, and B. Van Waeyenberge, The design and verification of MuMax3, AIP advances **4**, 107133 (2014).

[41] L. Caretta, M. Mann, F. Büttner, K. Ueda, B. Pfau, C. M. Günther, P. Hessing, A. Churikova, C. Klose, and M. Schneider, Fast current-driven domain walls and small skyrmions in a compensated ferrimagnet, Nature nanotechnology **13**, 1154 (2018).

[42] L. Herrera Diez, F. Ummelen, V. Jeudy, G. Durin, L. Lopez-Diaz, R. Diaz-Pardo, A. Casiraghi, G. Agnus, D. Bouville, and J. Langer, Magnetic domain wall curvature induced by wire edge pinning, Applied Physics Letters **117**, 062406 (2020).

[43] A. Krasheninnikov and K. Nordlund, Ion and electron irradiation-induced effects in nanostructured materials, Journal of applied physics **107**, 3 (2010).

[44] I. Shorubalko, K. Choi, M. Stiefel, and H. G. Park, Ion beam profiling from the interaction with a freestanding 2D layer, Beilstein journal of nanotechnology **8**, 682 (2017).

[45] P.-H. Lin, B.-Y. Yang, M.-H. Tsai, P.-C. Chen, K.-F. Huang, H.-H. Lin, and C.-H. Lai, Manipulating exchange bias by spin–orbit torque, Nature Materials **18**, 335 (2019).

[46] Z. Liu, Z. Luo, I. Shorubalko, C. Vockenhuber, L. J. Heyderman, P. Gambardella, and A. Hrabec, Dataset for Strong lateral exchange coupling and current-induced switching in single-layer ferrimagnetic films with patterned compensation temperature, Zenodo.

**Supplementary Note 1: Sample fabrication**

The magnetic stack of Ta (1 nm)|Pt (5 nm)|GdCo (*x* nm)|Ta (2 nm) is deposited via magnetron sputtering on $Si_3N_4$|Si substrates at a base pressure below $3 \times 10^{-8}$ Torr and an Ar sputtering pressure of 3 mTorr using a commercial sputtering system. We deposited the GdCo layer by co-sputtering from Gd and a Co targets. We varied the thickness of the GdCo layer while keeping the atomic Co fraction of 69.9% fixed. This allows us to modify the compensation temperature $T_M$ of the ferrimagnetic GdCo layer (see Supplementary Note 2).

The oxidation process is performed in a commercial Oxford plasma chamber with an oxygen plasma power between 30 and 40 W for a time period of 60 – 90 s. The oxidation of the layer has been confirmed quantitatively by measuring the oxygen concentration in non-oxidized and oxidized samples via 13 MeV [127]I heavy ion elastic recoil detection analysis (ERDA). Example depth profiles for O are shown in Supplementary Fig. 1. Although, the sample thickness is slightly below the depth resolution of the technique, there is clear evidence that the total oxygen content in the Ta capping layer and ferrimagnetic CoGd layer increases by 40% after the oxygen plasma treatment, while the profiles of the other elements are not influenced.

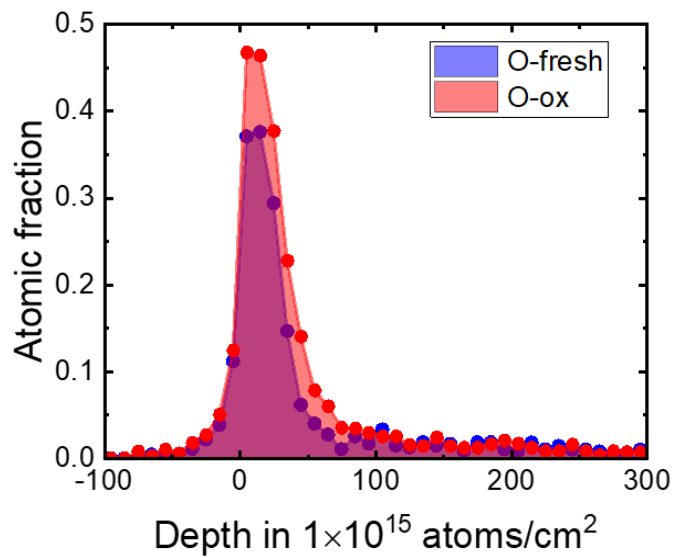

**Supplementary Figure 1:** 13 MeV [127]I heavy ion ERDA measurement to determine the difference in the atomic fraction of oxygen between Ta|Pt|CoGd|Ta thin films before (fresh) and after (ox) oxygen plasma treatment.

The designed patterns are prepared by electron beam lithography (EBL). We use 330 nm thick 4% PMMA in 950K molecular weight ethyl lactate as an electron-beam resist. The same resist is also used as a protection mask for the oxidation process.

**Supplementary Note 2: GdCo $T_M$ thickness dependence**

By changing the thickness of the GdCo layer, the compensation temperature of the GdCo film can be tuned. The change in $T_M$ with thickness is a result of the variation in the composition across the film thickness due to the formation of microstructures such as island and voids [38,47]. As shown in Supplementary Fig. 2, by decreasing the thickness of GdCo layer from 6.3 nm to 3.3 nm, a reversal of the hysteresis loop measured by polar-MOKE at room temperature is observed, which indicates that $T_M$ changes from being above 300 K to being below 300 K. This is also accompanied by divergence in the coercive field at the thickness where the sample is magnetically compensated at room temperature.

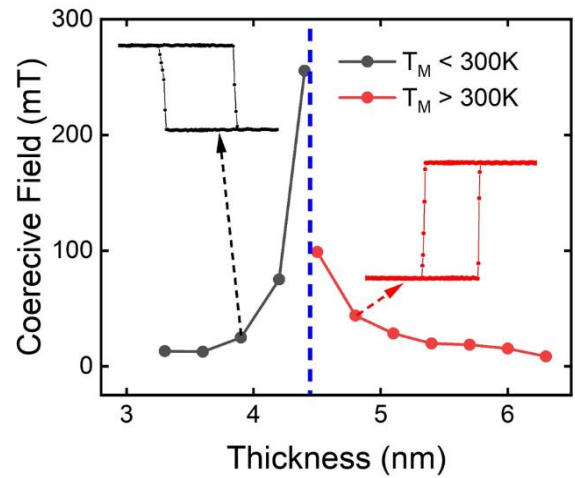

**Supplementary Figure 2:** Room temperature coercive fields measured using polar-MOKE for GdCo films with variable thickness. The insets depict two representative hysteresis loops taken for thicknesses smaller (in black) and larger (in red) than the thickness at which the magnetic film is compensated at room temperature.

**Supplementary Note 3: Effect of oxidation on the $T_M$ of the GdCo films**

By introducing oxygen into the GdCo layer, we can reduce the $T_M$ of the original film. As shown in Supplementary Fig. 3, the $T_M$ of the as-grown film of 4.6 nm thick GdCo changes from ≈260 K to ≈190 K on oxidation. The hysteresis loops, with the coercive field diverging at the compensation temperatures, are recorded via Hall resistance measurement in 1 μm Hall bar devices.

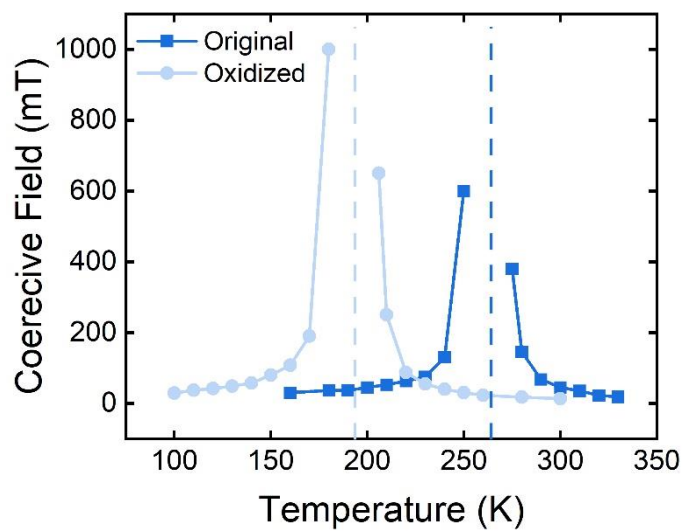

**Supplementary Figure 3:** Anomalous Hall effect measurement of the coercivity of GdCo (4.6 nm) before (dark blue) and after (light blue) oxidation.

Similarly, by oxidizing a 6.3 nm thick GdCo film whose $T_M$ is higher than room temperature, we could bring the $T_M$ of the film below room temperature. This scenario can be verified by polar-MOKE measurement at room temperature as shown in Supplementary Fig. 4, where the sign of the hysteresis loops is reversed.

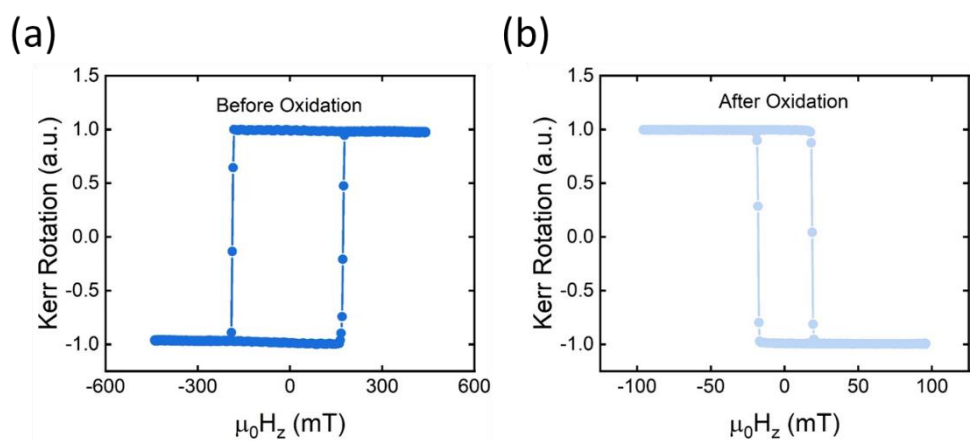

**Supplementary Figure 4:** Polar MOKE hysteresis loop measurement of a GdCo (6.3 nm) film at room temperature (a) before and (b) after oxidation.

The film topography is also affected by the oxygen absorption. As shown in Supplementary Fig. 5, the height profile along the blue line in the atomic force micrograph reveals a 1 nm increase in the thickness after the oxidation.

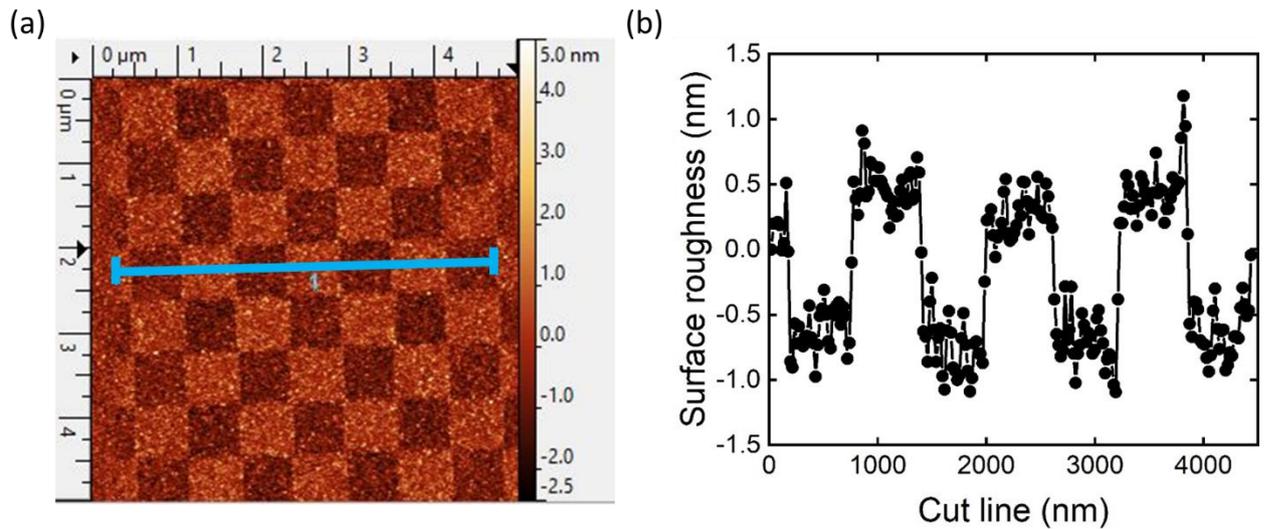

**Supplementary Figure 5** (a) Atomic force micrograph of the checkerboard pattern in GdCo (6.3 nm). The bright and dark regions correspond to the measured height of oxidized and non-oxidized regions, respectively. (b) Height profile along the line shown in panel (a) reveals a height difference of 1 nm between oxidized and non-oxidized regions.

**Supplementary Note 4: Transport and wide-field Kerr measurements**

The anomalous Hall electrical measurements are performed in a commercial Physical Property Measurement System (PPMS). The samples are measured with a sensing current of 100 µA under standard DC drive mode with an average of 10 readings per data point.

The Kerr images are recorded using a commercial wide-field Kerr microscope in the polar configuration. The sequence of Kerr images in Fig. 3(a) are captured while continuously decreasing the field to zero at 0.5 mT/sec. The magnetic contrast is visualized using differential Kerr imaging where the images are obtained by subtraction from the background image. The background image used in Fig. 3 was taken at 5 mT, while the background images in Fig. 1 and Fig. 4 are captured at 0 mT after applying field of 250 mT.

**Supplementary Note 5: Domain wall driven exchange bias**

For the 50 µm × 50 µm square design and the corresponding exchange bias loop measurements shown in Fig. 1(g), the switching is due to interfacial exchange coupling induced domain nucleation and corresponding domain wall motion as shown in Supplementary Fig. 6.

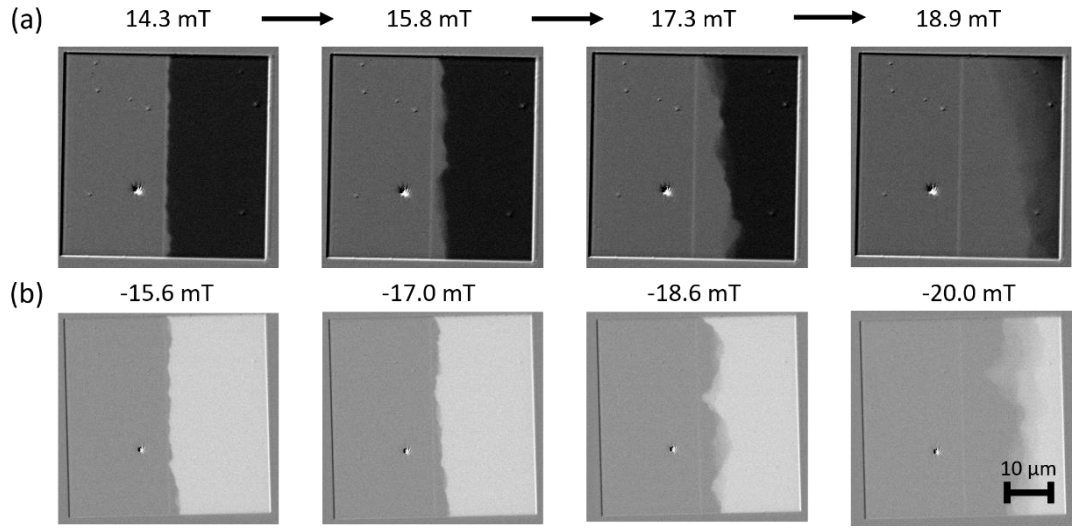

**Supplementary Figure 6:** A series of Kerr images of GdCo(6.3 nm) recorded during the exchange bias measurement [see Fig. 1(g) of the main text], (a) with negative exchange bias and (b) with positive exchange bias. The straight vertical line in the middle of the square separates the oxidized (right) and non-oxidized (left) regions.

**Supplementary Note 6: Anomalous Hall effect measurements of the lateral exchange coupling field**

Anomalous Hall effect (AHE) measurements of selectively oxidized GdCo(4.6 nm) tracks (see Fig. 2a of the main text) reveal a composite hysteresis loop consisting of a narrow square loop centered around zero field (larger signal from surrounding non-oxidized region) and two side loops at higher field (smaller signal from oxidized track region). The side loops arise from the lateral exchange coupling between the regions with $T_M$ larger and smaller than the measurement temperature (ranging from 220K to 260K). Depending on the temperature, we observe two types of loops: as shown in Supplementary Fig. 7: (a) the exchange coupling field minor loops are separated from the central hysteresis loop; (b) the exchange coupling field minor loops are merged with the central hysteresis loop. The shape of the loops depends on the relative strength of exchange coupling field between the non-oxidized region and the oxidized track as we change the temperature.

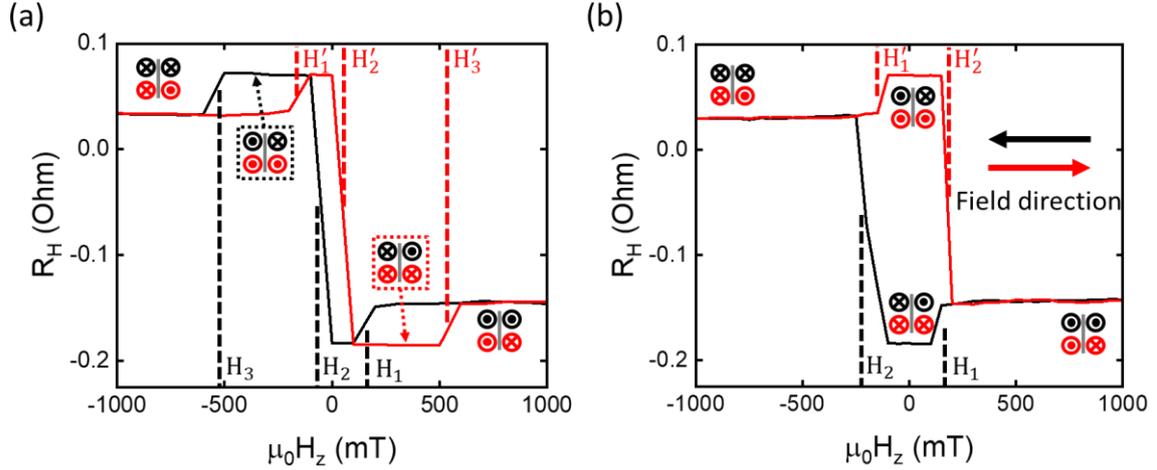

**Supplementary Figure 7:** Full hysteresis loops of a 200-nm-wide oxidized track in CoGd (4.6 nm) at 230 K (a) and 240 K (b). The black and red lines correspond to the field sweep direction from +1 T to -1 T and *vice versa*, respectively. The relevant magnetization states of oxidized|non-oxidized regions are depicted (black for $\boldsymbol{M}_{\text{net}}$ and red for $\boldsymbol{M}_{\text{co}}$).

In Supplementary Fig. 7(a) is shown the full hysteresis loop measured via Hall resistance as a function of OOP magnetic field ranging from +1T to -1T (black line) and back (red line). The loops are recorded at 230 K using a Hall bar as shown in the inset of Fig. 2(a) in the main text. At a field of +1 T, the net magnetization of the oxidized track and the surrounding non-oxidized region are parallel to each other. The AHE resistance has an intermediate value because it reflects the overall antiparallel alignment of the Co sublattice magnetization in the two regions. On reducing the applied field from +1 T to $H_1$, the interfacial exchange coupling overcomes the Zeeman energy. Considering that the non-oxidized region is much larger than the oxidized region, only the oxidized DW track can be switched by the interfacial exchange coupling via a DW propagation mechanism. Thus, the net magnetization of the oxidized track switches to the exchange-favored state in which it is antiparallel to that of the surrounding region. This configuration corresponds to the maximum amplitude of the AHE resistance because of the parallel alignment of the Co magnetic moments. At $H_2$, the magnetization of the surrounding region and that of the oxidized track switch simultaneously due to application of a reversed magnetic field that exceeds the coercivity of the non-oxidized region. Finally, once the Zeeman energy again overcomes the exchange coupling energy at $H_3$, the magnetization of the oxidized track switches so that its net magnetization follows the applied magnetic field direction. This corresponds to the net magnetization (Co magnetic moments) inside and surrounding the oxidized track being aligned parallel (antiparallel) to each other, analogous to the initial configuration at +1 T but with opposite magnetization direction. The field $H'_1$, $H'_2$ and $H'_3$ marked in Supplementary Fig. 7(a) refer to the similar situation when the field is swept from -1 T to +1 T (red line). The exchange coupling field is defined as the field value at the center of the minor hysteresis loop, which is calculated as $H_{\text{EC}} = \frac{|H_1 + H'_3| + |H_3 + H'_1|}{4}$. In Supplementary Fig. 7(b) is shown the hysteresis loop of the same sample recorded at 240 K. In this case, the interfacial exchange coupling strength is smaller than the coercivities of the surrounding region $H_2$ and $H_2'$. Thus, the minor loops merge into the central loop, and $H_3$ and $H_3'$ are not distinguishable anymore.

Moreover, the relative signal arising from the oxidized track depends on its width. As shown in Supplementary Fig. 8, given the same Hall bar width of 1 μm, with different track widths of 200 nm and 100 nm, the minor loop signal height of the 100 nm track is roughly half the signal height of the 200 nm track. This further demonstrates that the minor loops are associated with the oxidized tracks.

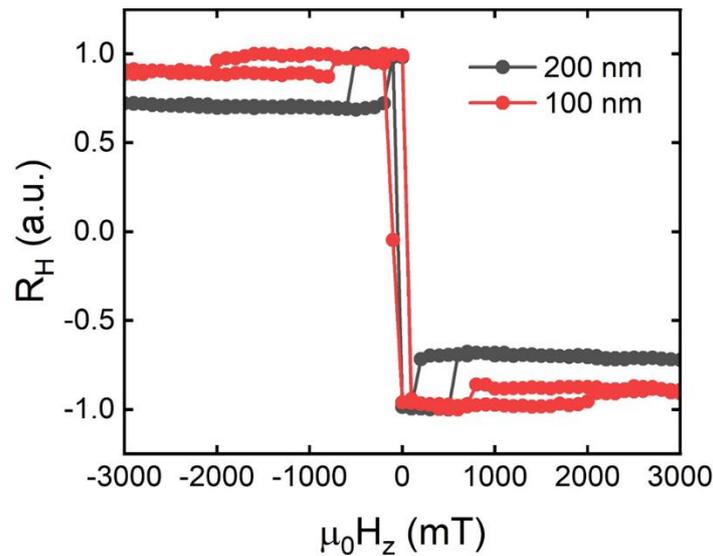

**Supplementary Figure 8:** Normalized full hysteresis loop measurement of 200-nm and 100-nm-wide oxidized tracks at 230 K. The height of the minor loop scales with the size of the oxidized track

**Supplementary Note 7: Details of FIB He⁺ irradiation**

The He⁺ irradiation is performed with a Focused Ion Beam (FIB) using a He⁺ microscope with a pattern generator. Cr (5 nm)|Au (20 nm) alignment markers are patterned on the pristine GdCo films using electron-beam lithography followed by lift-off. Then the track patterns are irradiated with a He⁺ current with doses ranging from $3 \times 10^{15}$ He⁺/cm² to $24 \times 10^{15}$ He⁺/cm². The He⁺ beam can produce features down to 5-10 nm and a 100-500 nm implantation depth at 30kV acceleration voltage (5-20 pA current) [44,48]. On injecting He⁺ ions into the GdCo films, the magnetic properties such as the anisotropy and the $T_M$ of the ferrimagnetic material are modified. These changes are due to vacancies induced by ion penetration, which modify the chemical short-range order, and an increase of oxygen atoms diffusing into the film that cause preferential oxidation of the Gd atoms [31].

The stopping fields are measurable only in the dose range 9 to $18 \times 10^{15}$ He$^+$/cm$^2$ [Supplementary Fig. 9(a)]. While doses smaller than $9 \times 10^{15}$ He$^+$/cm$^2$ do not bring the compensation temperature of the irradiated region below room temperature, doses higher than $20 \times 10^{15}$ He$^+$/cm$^2$ result in sample damage. Within the irradiation dose range that results in lateral exchange coupling, a decrease in the stopping field is observed when increasing the He$^+$ dose. Different track widths ranging from 200 nm to 2 µm are irradiated with a dose of $9 \times 10^{15}$ He$^+$/cm$^2$. The corresponding stopping fields shown in Supplementary Fig. 9(b) increase as the track width is reduced, which reflects the interfacial origin of the coupling effect. The He$^+$ irradiation technique therefore offers an alternative patterning method to oxidation with exquisite control over the desired coupling strength by changing the irradiation dose.

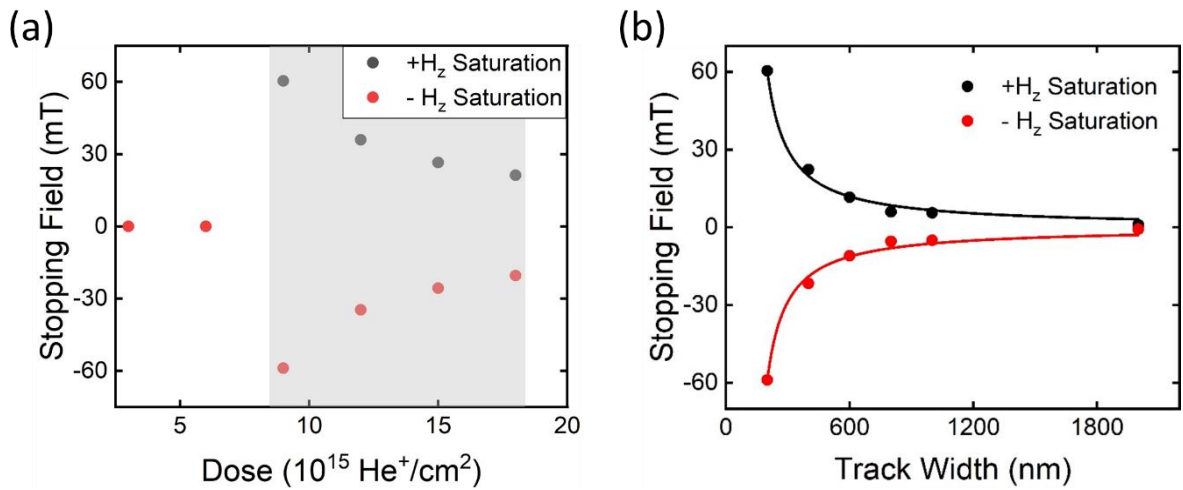

**Supplementary Figure 9:** (a) Stopping field *versus* irradiation dose in 6.5 nm GdCo patterned by He$^+$ irradiation. The gray shading indicates the dose range where the stopping field is measurable. (b) Stopping field *versus* track width in 6.5 nm GdCo patterned by He$^+$ irradiation. The lines in (b) are fits according to Eq. (1) in the main text. The thickness of GdCo is 6.5 nm.

**Supplementary Note 8: Micromagnetic simulations of the effective coupling field with and without dipolar field**

To evaluate the effect of the dipolar field on the ground state, a 2D micromagnetic simulation is performed via Mumax$^3$ [40]. The material parameters used in the simulation are as follows: $D = 0.2 \times 10^{-3}$ J/m$^2$, the saturation magnetization is $8.2 \times 10^4$ A/m in the non-oxidized region and $2.3 \times 10^5$ A/m in the oxidized region, as obtained from SQUID measurements. The exchange stiffness is $A_\text{ex} = 7 \times 10^{-12}$ J/m and OOP uniaxial anisotropy constant $K_\text{u} = 86$ kJ/m$^3$. We also considered periodic boundary conditions. To quantify the magnitude of the effective OOP magnetic field acting on the central stripe, we determined the energies of ⊙⊙⊙ and ⊙⊗⊙ states ( $E_{⊙⊙⊙}$ and $E_{⊙⊗⊙}$ ) from 1D micromagnetic simulations [18]. As sketched in the inset of Supplementary Fig. 10, the central oxidized track is antiferromagnetically coupled to the surrounding film via two cells with a negative exchange coupling ($J_{AP}$) at both interface. By performing systematic simulations, we find that the dipolar field contributes marginally to the effective field. We conclude that the coupling is mostly mediated by the proposed lateral exchange coupling mechanism.

The simulations also show that $J_{AP}$ is of the order of 10% of the exchange coupling $A_{ex}$. In this simplified model, a single homogenous exchange parameter between the Co-Co magnetic moments over the whole GdCo film is considered. While the Co-Gd coupling is strong enough to maintain ferrimagnetic coupling, the Gd-Gd coupling strength is much smaller than between Co-Co magnetic atoms. The coupling strength estimated from the simulations (see Eq. 1 of the main text) is $J_{AP} \approx 0.7 \times 10^{-12}$ J/m [41]. This number can be compared to the analytically estimated number, which is $J_{AP} = 4\sqrt{AK_{eff}} - \pi D \approx 16.1 \times 10^{-12}$ J/m. The estimated numbers are susceptible to the uncertainty in the value of DMI and magnetic anisotropies used in the simulation, which affect the DW energy but not the effective fields in micromagnetic simulations. The computed effective fields are also in a relatively good agreement with the measured stopping fields presented in Fig. 3, which are replotted in Supplementary Fig. 10.

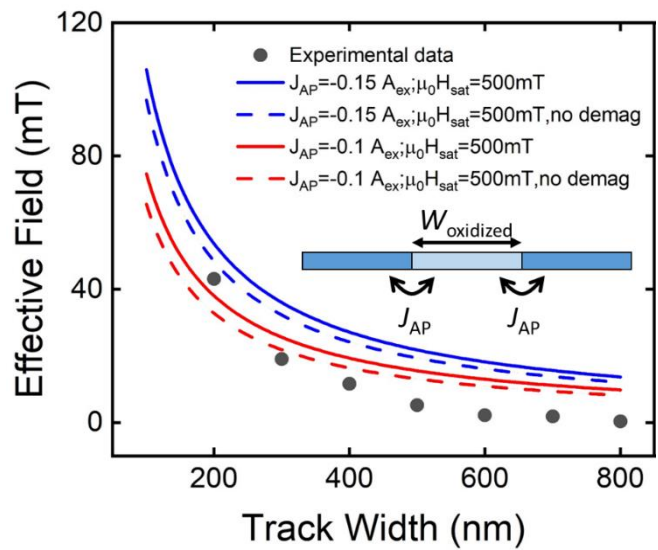

**Supplementary Figure 10:** Experimental data (dots) and simulated effective field versus track width in the presence (full lines) or absence (dashed lines) of dipolar field. A schematic of the simulated geometry is shown in the inset.

**Supplementary Note 9: Reproducible electrical switching of exchange bias**

To illustrate the statistical significance of the data presented in Fig. 4, we probed the repeated reversal of exchange bias by the application of current pulses of fixed polarity while reversing the polarity of the longitudinal in-plane magnetic field. A highly-reproducible switching of exchange bias can be observed in Supplementary Fig. 11.

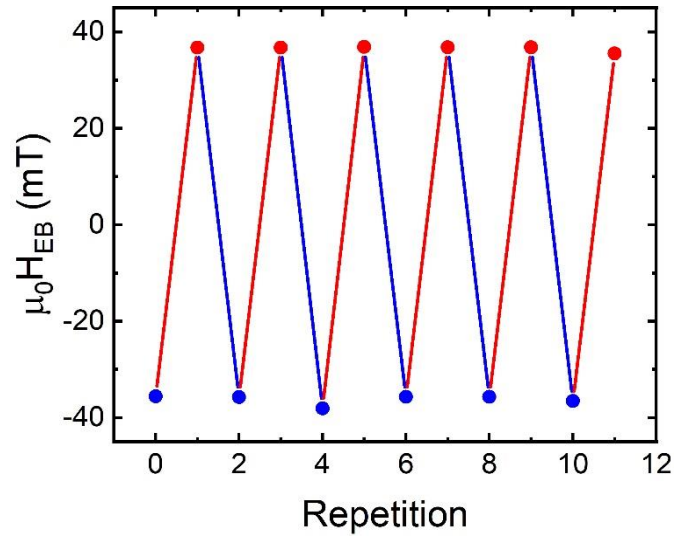

**Supplementary Figure 11:** Exchange bias measurement of cross-shape structure after applying $50 \times 150$-ns-long current pulses at a current density of $1.38 \times 10^{11}$ A/m² while alternating an in-plane magnetic field between $-250$ mT (red) and $+250$ mT (blue) applied along the current direction.

**Reference:**

[1] R. Malmhäll and T. Chen, Thickness dependence of magnetic hysteretic properties of rf‐sputtered amorphous Tb–Fe alloy thin films, Journal of Applied Physics **53**, 7843 (1982).

[2] B. Hebler, A. Hassdenteufel, P. Reinhardt, H. Karl, and M. Albrecht, Ferrimagnetic Tb–Fe Alloy thin films: composition and thickness dependence of magnetic properties and all-optical switching, Frontiers in Materials **3**, 8 (2016).

[3] I. Shorubalko, K. Choi, M. Stiefel, and H. G. Park, Ion beam profiling from the interaction with a freestanding 2D layer, Beilstein journal of nanotechnology **8**, 682 (2017).

[4] I. Shorubalko, L. Pillatsch, and I. Utke, in *Helium Ion Microscopy* (Springer, 2016), pp. 355.

[5] M. Krupinski, J. Hintermayr, P. Sobieszczyk, and M. Albrecht, Control of magnetic properties in ferrimagnetic GdFe and TbFe thin films by He+ and Ne+ irradiation, Physical Review Materials **5**, 024405 (2021).

[6] A. Vansteenkiste, J. Leliaert, M. Dvornik, M. Helsen, F. Garcia-Sanchez, and B. Van Waeyenberge, The design and verification of MuMax3, AIP advances **4**, 107133 (2014).

[7] Z. Liu, Z. Luo, S. Rohart, L. J. Heyderman, P. Gambardella, and A. Hrabec, Engineering of Intrinsic Chiral Torques in Magnetic Thin Films Based on the Dzyaloshinskii-Moriya Interaction, Physical Review Applied **16**, 054049 (2021).

[8] L. Caretta, M. Mann, F. Büttner, K. Ueda, B. Pfau, C. M. Günther, P. Hessing, A. Churikova, C. Klose, and M. Schneider, Fast current-driven domain walls and small skyrmions in a compensated ferrimagnet, Nature nanotechnology **13**, 1154 (2018).